\documentclass[%
reprint,%
 amssymb, amsmath,%
 aip,apl,%
]{revtex4-1}
\usepackage{docs}%

\usepackage{bm}%
\usepackage[colorlinks=true,linkcolor=blue]{hyperref}%
\usepackage{epsfig}
\expandafter\ifx\csname package@font\endcsname\relax\else
 \expandafter\expandafter
 \expandafter\usepackage
 \expandafter\expandafter
 \expandafter{\csname package@font\endcsname}%
\fi
\begin{document}
\title{Backward magnetostatic surface spin waves in exchange coupled Co/FeNi bilayers}%
\author{Wenjie Song}
\altaffiliation[]{These authors contributed equally to this work.}
\affiliation{Key Laboratory for Magnetism and Magnetic Materials of the Ministry of Education, Lanzhou University, Lanzhou, 730000, People's Republic of China}%
\author{Xiansi Wang}
\altaffiliation[]{These authors contributed equally to this work.}
\affiliation{Center for Quantum Spintronics, Department of Physics,
		Norwegian University of Science and Technology, NO-7491 Trondheim, Norway}%

\author{Wenfeng Wang}
\affiliation{Key Laboratory for Magnetism and Magnetic Materials of the Ministry of Education, Lanzhou University, Lanzhou, 730000, People's Republic of China}%
\author{Changjun Jiang}
\affiliation{Key Laboratory for Magnetism and Magnetic Materials of the Ministry of Education, Lanzhou University, Lanzhou, 730000, People's Republic of China}%

\author{Xiangrong Wang}
\email{phxwan@ust.hk}
\affiliation{Department of Physics, The Hong Kong University of Science and Technology, Clear Water Bay, Kowloon, Hong Kong, People's Republic of China}%
\affiliation{HKUST Shenzhen Research Institute, Shenzhen 518057, People's Republic of China}%
\author{Guozhi Chai}
\email{chaigzh@lzu.edu.cn}
\affiliation{Key Laboratory for Magnetism and Magnetic Materials of the Ministry of Education, Lanzhou University, Lanzhou, 730000, People's Republic of China}%
\date{\today}
\begin{abstract}
Propagation of backward magnetostatic surface spin waves (SWs) in exchange coupled Co/FeNi bilayers are studied by using Brillouin light scattering (BLS) technique. Two types of SWs modes were identified in our BLS measurements. They are magnetostatic surface waves (MSSWs) mode and perpendicular standing spin waves (PSSWs) mode. The dispersion relations of MSSWs obtained from the Stokes and Anti-Stokes measurements display respectively positive and negative group velocities. The Anti-Stokes branch with positive phase velocities and negative group velocities, known as backward magnetostatic surface mode originates from the magnetostatic interaction of the bilayer. The experimental data are in good agreement with the theoretical calculations. Our results are useful for understanding the SWs propagation and miniaturizing SWs storage devices.
\end{abstract}



\maketitle



Spin waves (SWs) in magnetic films have attracted much attention in recent years because of great potential applications in spintronic devices.\cite{1,2,3,4,5} SWs are collective oscillations of gigahertz frequency in typical ferromagnetic materials.\cite{15,16} The SW wavelength is orders of magnitude shorter than that of electromagnetic waves of the same frequency, so that they can be used in micro or nano size spin wave devices.\cite{2,3,15,16} SWs can be classified as exchange type and magnetostatic one according to the dominating interactions.\cite{19} It is well-known that there are three different kinds of magnetostatic SWs, including the magnetostatic surface waves (MSSWs) which is also called Damon-Eshbach (DE) mode, the backward volume magnetostatic spin waves (BVMSWs) and the forward volume magnetostatic spin waves (FVMSWs).\cite{3,21,22,23,24} The MSSWs are nonreciprocal and exhibit different characteristics at different interfaces of the films.\cite{3,22} For both MSSWs and FVMSWs in single layer films, their dispersion relations give both positive phase and group velocities with, in general, different magnitudes.\cite{3,22} The dispersion relation of BVMSWs has positive phase velocities and negative group velocities. This backward property could be useful in velocity-related applications such as the inverse Doppler effect, etc.\cite{3,29,30} One interesting question is whether a negative group velocity can arise in MSSWs in a hybridized system. Such backward magnetostatic surface wave (BMSSWs) should be very interesting and important because MSSWs are useful in magnonics as information carriers for realizing SW-based devices, such as SW filters, SW beamspliters, and SW emitters.\cite{1,33,34,35,36}

Nonreciprocal property of MSSWs says that a MSSW propagates along $\textbf{m}\times \textbf{n}$. Here \textbf{m} is the magnetization direction and \textbf{n} is the normal direction of the surface that points outward.\cite{21} If two magnetic layers are stacked together, two sets of MSSWs propagating in opposite direction with each other shall couple each other. Previous works on very thin bilayers has shown that coupling between two counter-propagating MSSWs can cause non-reciprocity,\cite{37,38,39,40,41} but not backward MSSWs. We suspect that backward MSSWs are possible when two counter-propagating MSSWs are sufficiently different from each other and their coupling are strong enough. The reason that backward MSSWs were not found before (Ref. 19-21) is either because the film thickness is too small so that the coupling of MSSWs on different surfaces is important or because two MSSWs are too similar. Thus, we will use thicker bilayers (Co/FeNi) in this study. Besides the usual nonreciprocal property of MSSWs from the magnetostatic interactions, the Dzyaloshinskii-Moriya interaction (DMI) is chiral in nature and can also lead to asymmetric dispersion relation that, in turn, results in nonreciprocal behavior of SWs.\cite{42,43,44} In this letter, we use Brillouin light scattering (BLS) to obtain dispersion relations of surface SWs of Co/FeNi bilayer system. These two materials are chosen because of their significant distinct saturation magnetizations and other magnetic properties. The MSSWs and perpendicular standing spin waves (PSSWs) were observed. The dispersion relation of MSSWs obtained from the Anti-Stokes measurements confirms that they are of BMSSWs while that obtained from Stokes measurements has the usual MSSW spectrum. The experimental results were also verified by numerical calculations with the material parameters.

Our Co(30)/FeNi($t$) bilayer films were deposited on single-crystal Si (111) substrates by radio frequency (RF) magnetron sputtering. The numbers in parentheses are the film thicknesses in nanometers. The FeNi layer thickness $t$ varied from 30 to 50 nm that was controlled by varying the sputtering time. The base pressure of sputtering chamber was about $5\times 10^{-5}$ Pa. The pressure inside the chamber was 0.3 Pa and the RF power was 50 W during sputtering process. Static magnetization of Co(30)/FeNi($t$) bilayer were measured by a vibrating sample magnetometer (VSM). Figure 1 shows of the in-plane magnetic hysteresis loops of bilayer samples at room temperature when the field is applied along the easy-axis (a) and along the hard-axis (b). As shown in Fig. 1, in both cases the magnetic hysteresis loops show smooth curves. This means that the bilayer are coupled like a single layer.\cite{53} FMR measurements and relevant experimental parameters can be found in supplementary materials.

\begin{figure}
\begin{center}
\epsfig{file=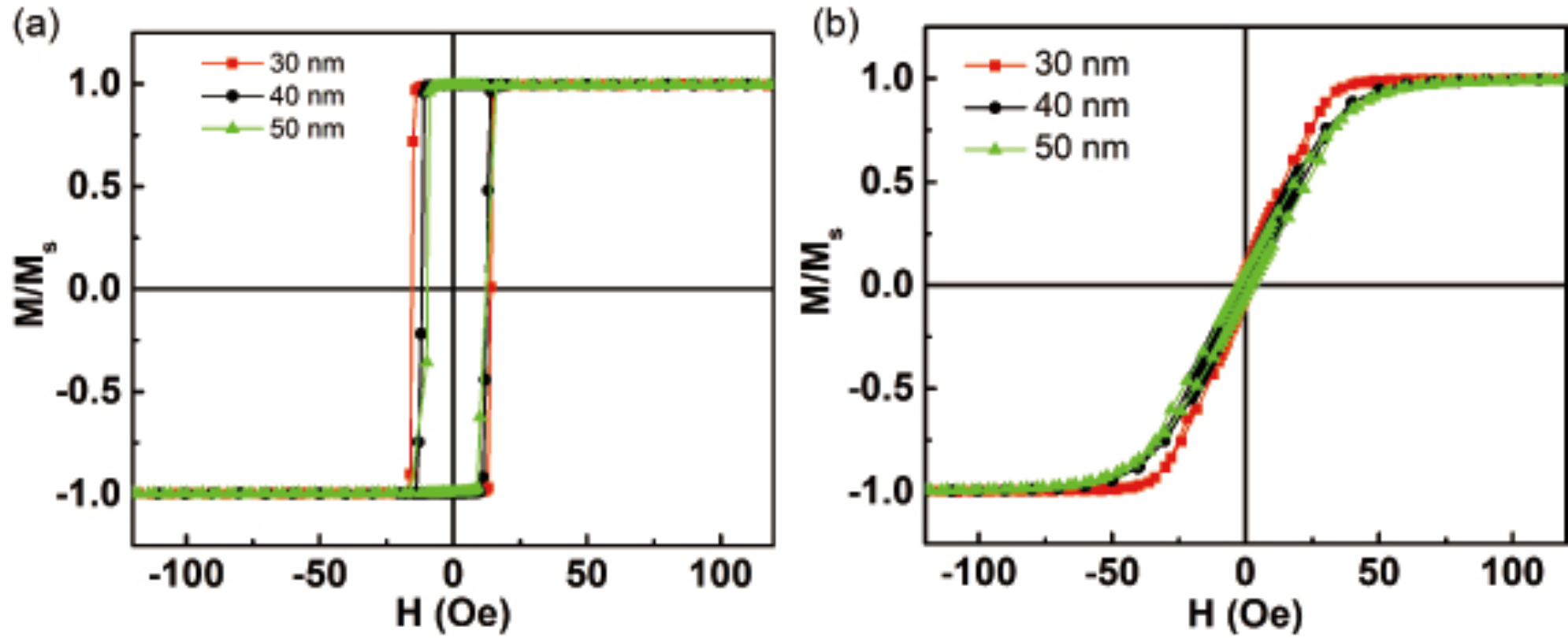,width=8.5 cm} \caption{
The normalized in-plane hysteresis loops of Co/FeNi bilayer. (a) The hysteresis loops of FeNi layer of various thicknesses. The field is along the easy-axis. (b) The hysteresis loops of FeNi layer of various thicknesses. The field is along the hard-axis.
}\label{Fig.1}
\end{center}
\end{figure}

BLS measurements were performed for the Co/FeNi bilayer at room temperature. BLS is very effective for achieving the vector resolution of surface waves.\cite{55,56,57} The scattering process can be described by inelastic scattering and is known as the $180^{\circ}$ backscattering geometry (see inset of Fig. 2(a)). In experiments, external magnetic field \textbf{H} is in-plane and perpendicular to the MSSWs wave vector. The incident plane of the laser light is perpendicular to the external field as shown in the inset of Fig. 2(a). The wave vector of the MSSWs is $k_{\parallel}$ = $4\pi {\rm sin}{\theta}/{\lambda}$, where the ${\theta}$ is the incident angle of the light and the ${\lambda}$ is the laser wavelength (532 nm). The laser power incident on the surface of the sample is about 30 mW in the experiments. In our work, the range of $k_{\parallel}$ varies from -16.53 rad/${\mu}$m to 16.53 rad/${\mu}$m with a step size of 1.18 rad/${\mu}$m by varying the laser light incident angle. The resolution of the scanning frequency is 0.068 GHz. The PSSWs wave vector does not change with $k_{\parallel}$. The typical BLS spectra for FeNi(30) single layer and Co(30)/FeNi(30) bilayer are illustrated in Fig. 2 with the absolute value of wave vector $k_{\parallel}$ = 16.53 rad/${\mu}$m and external magnetic field \textbf{H} = 500 Oe. In the light scattering process, the direction of Anti-Stokes (positive frequency) is defined as the direction of positive wave vector and the direction of Stokes (negative frequency) is defined as the direction of negative wave vector. From Fig. 2, both the FeNi(30) single layer and the Co(30)/FeNi(30) bilayer spectrum display four different peaks. The two peaks of larger signal are for MSSWs and the other two peaks with weaker signal are for PSSWs. For the FeNi(30) single layer, the MSSWs show one Stokes peak $f(-k)$ with a negative frequency shift (-12.87 GHz) and one Anti-Stokes peak $f(k)$ with positive frequency shift (12.82 GHz). This suggests that the two MSSWs propagate in opposite directions and with the same absolute value of frequency within experimental error. However, for the Co(30)/FeNi(30) bilayer, the MSSWs of Co(30)/FeNi(30) bilayer show one Stokes peak $f(-k)$ with a negative frequency shift (-13.28 GHz) and one Anti-Stokes peak $f(k)$ with positive frequency shift (9.67 GHz). This suggests that the two MSSWs propagate in opposite directions and have a distinct frequency difference. In contrast, for the PSSWs mode, the frequency of both FeNi(30) single layer and Co(30)/FeNi(30) bilayer are almost the same.

\begin{figure}
\begin{center}
\epsfig{file=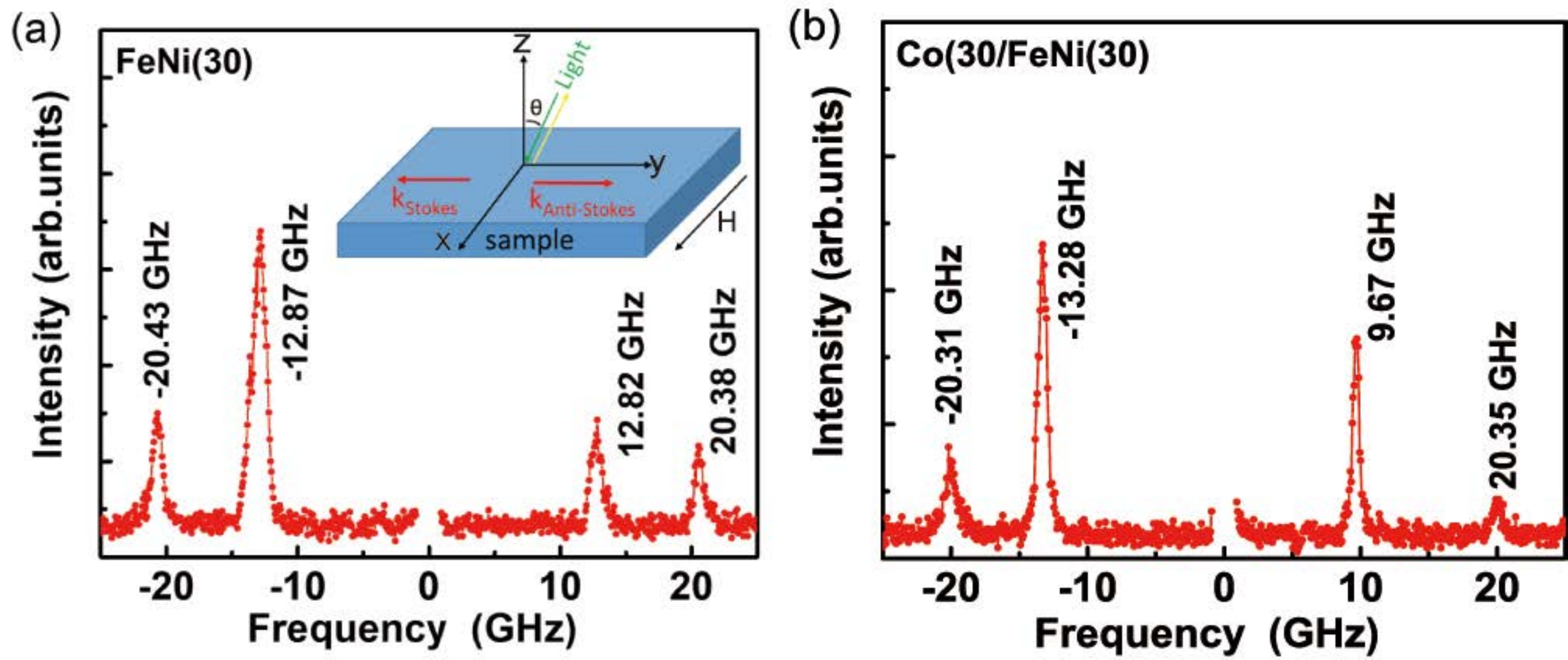,width=8.5 cm} \caption{
BLS spectrum measured for (a) FeNi(30) single layer and (b) Co (30)/FeNi(30) bilayer with the absolute value of wave vector $k_{\parallel}$ = 16.53 rad/${\mu}$m and external magnetic field \textbf{H} = 500 Oe. Inset: Schematic of the scattering process of $180^{\circ}$ backscattering geometry. The incident angle is denoted by ${\theta}$. Magnetic field H is along the $x$-direction.
}\label{Fig.2}
\end{center}
\end{figure}

\begin{figure*}
\begin{center}
\epsfig{file=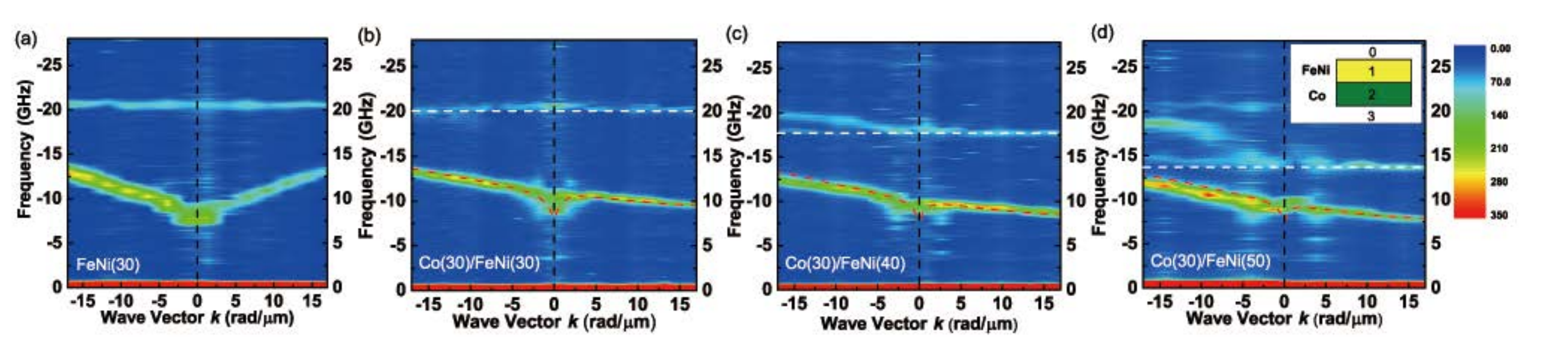,width=18 cm} \caption{
Density plot of BLS intensity in $\omega$-k plane for FeNi(30) single layer (a) and Co/FeNi bilayer with various FeNi layer thicknesses of (b) $t$ = 30 nm, (c) $t$ = 40 nm, (d) $t$ = 50 nm. The two sides of the vertical black dotted line correspond to different wave vector coordinates, respectively. The red dashed curves are of calculated MSSW spectrum. The white dashed curves are of the PSSW spectrum. The fixed external field $\textbf{H} = 500$ Oe is applied along the $x$-direction. Inset: Schematic diagram of the bilayer system.
}\label{Fig.3}
\end{center}
\end{figure*}

The dispersion relation of FeNi(30) single layer and Co(30)/FeNi($t$) bilayers with various FeNi thicknesses of $t=30,$ 40, and 50 nm was measured by BLS. Figure 3 is the density plot of BLS intensity in frequency - wave vector plane. The left side of the vertical black dotted line correspond to negative wave vector (Stokes) and the right side of the vertical black dotted line correspond to positive wave vector (Anti-Stokes). In the experiments, a constant magnetic of \textbf{H} = 500 Oe is along the $x$-direction and the $k_{\parallel}$ varies from -16.53 rad/${\mu}$m to 16.53 rad/${\mu}$m along the $y$-direction. As shown in Fig. 3(a), the lower frequency mode is MSSWs and higher frequency mode is PSSWs. For lower frequency mode, the frequency increases with the wave vector. For higher frequency mode, the frequency does not change with the wave vector.  As shown the Co(30)/FeNi($t$) bilayers in Fig. 3(b),(c),(d), the lower frequency mode is MSSWs and higher frequency mode is a mixture of PSSW and optical branch. For lower frequency mode, the $f(k-)$ increases with the increase of wave vector $k$, which behaves like the common MSSWs. However, the $f(k+)$ decreases with the increase of wave vector $k$, which behaves like the BMSSWs. The data of negative magnetic field (see supplementary materials) further prove the results of backward magnetostatic surface spin waves.

To further substantiate our results, we theoretically calculate the spectrum of MSSWs of Co/FeNi bilayer. The sample is schematically illustrated in the inset of Fig. 3(d). The sample is assumed to be infinite in $xy$-plane, and an magnetic field \textbf{H} is applied along $x$ direction, the same as the experimental setup. The four regions denoted by $0\sim3$ are half-infinite vacuum, FeNi of thickness $d_1$, Co of thickness $d_2$, and half-infinite vacuum, respectively. For each region, the magnetostatic Maxwell's equations can be written\cite{38,81}
\begin{equation}\label{2}
 \nabla\cdot{\textbf{B}_{i}}={\mu_0}\nabla\cdot({\textbf{H}_{i}}+{\textbf{M}_{i}})=0,
\end{equation}
\begin{equation}\label{3}
 \nabla\times{\textbf{H}_{i}}=0,
\end{equation}
where $|{\textbf{M}_{i}}|=0$ for vacuum $i=0,3,$ $|{\textbf{M}_{1}}|=M_{s,\rm FeNi},$ and $|{\textbf{M}_{2}}|=M_{s,\rm Co}.$ The Maxwell equation implies boundary conditions that $\textbf{\^{z}}\times\textbf{H}$ and $\textbf{\^{z}}\cdot\textbf{B}$ are continuous at boundaries 01, 12, 23. The relation between ${\textbf{H}_{i}}$ and ${\textbf{M}_{i}}$ in the magnetic layers 1 and 2 are written by the Landau-Lifshitz-Gilbert (LLG) equation as\cite{82}
\begin{equation}\label{4}
\frac{\partial \textbf{M}_i}{\partial t}=-\gamma{\textbf{M}_{i}}\times(\frac{A_i}{{\mu_0}{M_{si}^{\rm 2}}}\nabla^2{\textbf{M}_{i}}+{\textbf{H}_{i}})+\frac{\alpha_i}{{M_{si}}}{\textbf{M}_i}\times\frac{\partial \textbf{M}_i}{\partial t},
\end{equation}
where $A_i$ are the intralayer exchange interactions in FeNi and Co,and $\alpha_i$ is the damping. If we do not consider surface effects such as surface anisotropy or surface spin transfer torque,\cite{81} the LLG equation implies boundary conditions $\frac{\partial \textbf{M}_i}{\partial z}=0$ at boundaries 01 and 23, and the boundary condition at boundary 12 is\cite{83}
\begin{equation}\label{2}
\frac{1}{2}\frac{A_{12}}{M_{s1}M_{s2}}(\textbf{M}_{1+}\times\textbf{M}_{2-})+\frac{A_{1}}{M_{s1}^2}(\textbf{M}_{1}\times\frac{\partial \textbf{M}_1}{\partial {\rm z}})=0
\end{equation}
\begin{equation}\label{2}
\frac{1}{2}\frac{A_{12}}{M_{s1}M_{s2}}(\textbf{M}_{2}\times\textbf{M}_{1})+\frac{A_{1}}{M_{s2}^2}(\textbf{M}_{2}\times\frac{\partial \textbf{M}_2}{\partial {\rm z}})=0
\end{equation}
where $A_{12}$ is the interfacial exchange interaction in units of $\rm J/m^2$. We expand ${\textbf{M}_{i}}$ and ${\textbf{H}_{i}}$ around their equilibrium values,
\begin{equation}\label{2}
\textbf{M}_{1,2}=(M_{s{1,2}},0,0)+\textbf{m}_{1,2}
\end{equation}
\begin{equation}\label{2}
\textbf{H}_{i}=(H,0,0)+\textbf{h}_{i}
\end{equation}
and keep only linear terms in the small quantities $\textbf{m}_{1,2}$ and $\textbf{h}_{i}$. By assuming a harmonic form $\textbf{m}_{1,2},\textbf{h}_{i}\sim e^{i{(\omega{t}-ky)}}$ and applying the boundary conditions, we obtain a secular equation whose solution is the dispersion relation. We show the two lower frequency branches which are MSSWs in Fig. 3 by red dash curves on top of the experimental data. The parameters are $M_{s1}=10$ kOe, $M_{s2}=17$ kOe, $\gamma_1=29.4$ GHz/T, and $\gamma_2=32.2$ GHz/T, obtained by a separate FMR experiment. The exchange interactions are reasonable fits: $A_1=10$ pJ/m, $A_2=11$ pJ/m, and $A_{12}=20$ $\rm{mJ/m^2}$. The plot shows good agreement between theoretical and experimental results. The white dash lines in Fig. 3 are of the PSSW spectrum. The frequency of PSSWs mode does not change with the wave vector of surface waves but decreases with the thickness of FeNi. Although the two layers are coupled, PSSWs can only exist in the upper layer since the saturation magnetizations of the two films are different. The PSSWs mode and the modes near PSSWs are closely related to exchange modes. The calculation is complicated and special studies will be done in further study.

\begin{figure}
\begin{center}
\epsfig{file=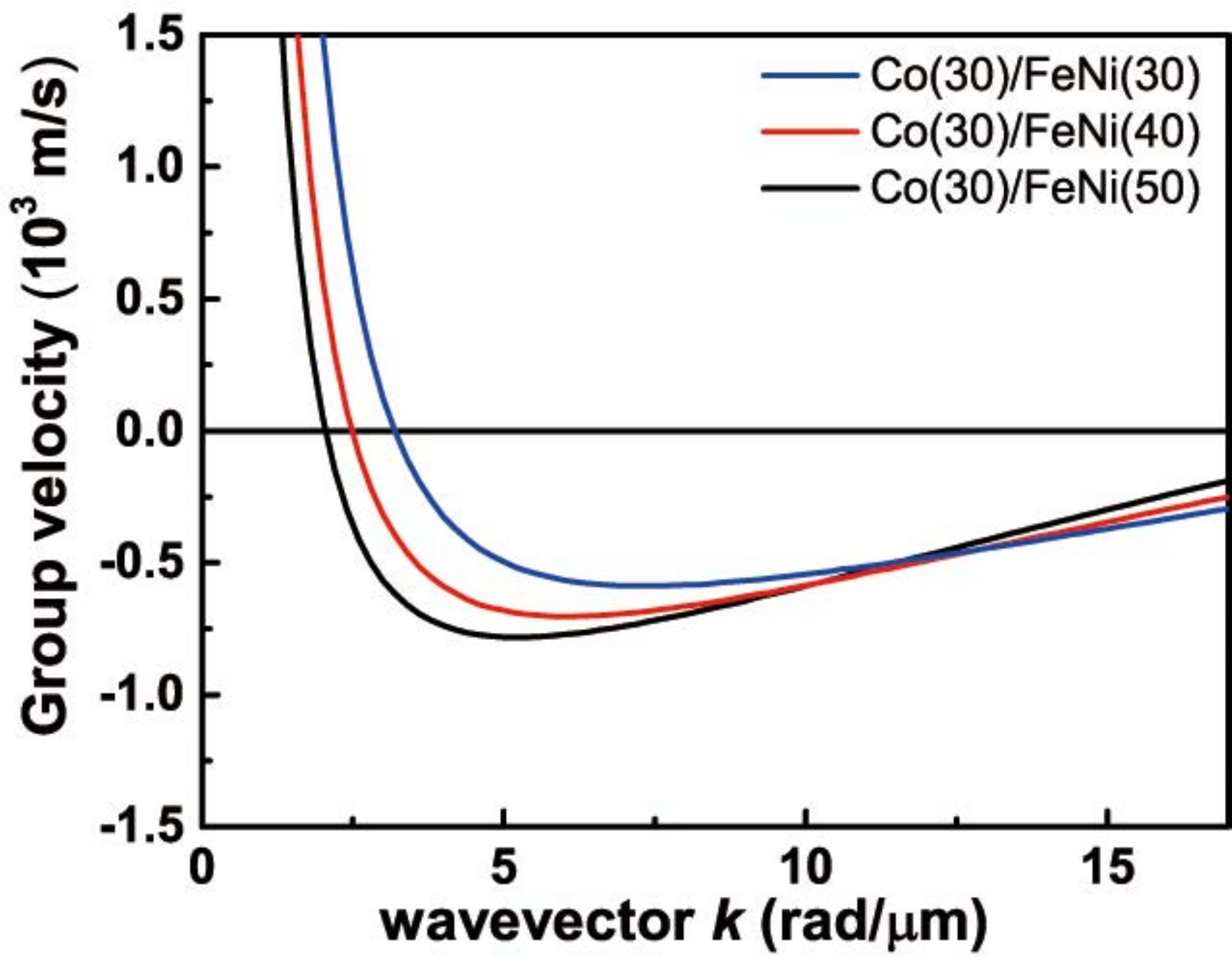,width=8.5 cm} \caption{
Theoretical calculations of group velocity as a function of positive wave vector for the different bilayer thicknesses.
}\label{Fig.4}
\end{center}
\end{figure}

Figure 4 shows theoretical calculations of group velocity as a function of positive wave vector with various model parameters. For different bilayer thicknesses, the phase velocities $\upsilon_p=\omega/k$ are always positive as shown in Fig. 3. While the group velocities $\upsilon_g=\partial \omega/\partial k$ are negative for most wave vector as shown in Fig. 4. For the FeNi layers of $t=30$, 40, and 50 nm, group velocities change sign from positive to negative at the wave vector of $k_{\parallel}$ = 3.17, 2.49, and 2.03 rad/${\mu}$m, respectively. The regions with opposite signs of phase velocities and group velocities are of BMSSWs. It occurs when two magnetic layers have very different saturation magnetizations. It is the magnetostatic interaction between the two layers that leads to nonreciprocal behavior of SWs and the BMSSWs.\cite{37,38,39,40,41} The largest magnitudes of negative group velocities for FeNi layers of $t = 30$, 40, and 50 nm, are respectively 0.59, 0.71, and $0.78\times10^3$ m/s. It can be interpreted that as the thickness increases, the exchange becomes weaker and the non-reciprocity becomes more obvious.\cite{39} So as the thickness of the FeNi layer increases, the negative slope is more steep.

In summary, we observed the BMSSWs in exchange coupled Co/FeNi bilayer by BLS. The results were further confirmed by theoretical calculations. We revealed that coupling of two counter propagating MSSWs at the interface, through the magnetostatic interaction between two layers, is responsible for the BMSSWs. Furthermore, the largest magnitudes of negative group velocities increase with the thinkness of FeNi.

\section*{SUPPLEMENTARY MATERIAL}

See the supplementary material for several experiments and detailed procedures of the derivation. While the main article contains the best representative data, other data are presented in the supplementary material. Part 1 is the ferromagnetic resonance(FMR) data and the relevant experimental parameters. Part 2 is the data of frequency varies with wave vector at magnetic field \textbf{H} = -500 Oe. Part 3 is the detailed procedures of the derivation of spin wave spectrums.

\begin{acknowledgments}
This work is supported by the National Natural Science Foundation of China (NSFC) (Nos. 51871117, 51471080, 51671099, 11804045 and 61734002), the Program for Changjiang Scholars and Innovative Research Team in University (No. IRT-16R35), and the Fundamental of Research Funds for the Central Universities: lzujbky-2018-118. X. R. W acknowledge supports from Hong Kong RGC (Grants No.16301518, 16301619 and 16300117 ).
\end{acknowledgments}

\end{document}